\documentstyle[aps,prl]{revtex}
\begin{document}
\title{Coulomb Gap: How a Metal Film Becomes an Insulator}
\draft
\author{V.Yu. Butko\cite{Vladimir}, J.F. DiTusa, and P.W. Adams}
\address{Department of Physics and Astronomy\\Louisiana State University\\Baton Rouge, Louisiana,
70806}
\date{\today}
\maketitle
\begin{abstract}
Electron tunneling measurements of the density of states (DOS) in
ultra-thin Be films reveal that a correlation gap mediates their insulating behavior.
In films with sheet resistance $R<5000\Omega$ the correlation singularity appears as the usual
perturbative $ln(V)$ zero bias anomaly (ZBA) in the DOS. As R is increased further, however, the
ZBA grows and begins to dominate the DOS spectrum.  This evolution continues until a
non-perturbative $|V|$ Efros-Shklovskii Coulomb gap spectrum finally emerges in the highest R
films.  Transport measurements of films which display this gap are
well described by a universal variable range hopping law $R(T)=(h/2e^2)exp(T_o/T)^{1/2}$.
	\\     
\end{abstract}

	It has been known for some time now that 2D is the lower critical dimension for disordered
transport and that even a non-interacting 2D electron gas will be localized in the
presence of arbitrarily small disorder in the thermodynamic limit
\cite{gang4,LeeRam}.  When these systems are probed at a finite length scale, by a
magnetic field for instance, logarithmic corrections to the Drude conductivity are
seen but one does not expect a true metal-insulator transition in 2D or even a
metallic phase in the conventional sense \onlinecite{LeeRam}.  What is somewhat less
well understood are the ramifications Coulomb interactions and their attendant
correlations, particularly in 2D systems with moderate to strong disorder.  This
issue in particular has now come under considerable scrutiny with the recent
discovery of an apparent metal-insulator transition in the dilute 2D electron
gas of Si metal-oxide-semiconductor field effect transistors (MOSFET's)
\cite{Krav95,Sim97}.  This somewhat surprising discovery runs so counter to
conventional wisdom that there has been speculation that perhaps $e-e$
interaction effects are stabilizing an anomalous 2D metallic phase \cite{Dobro}. 
Clearly, a systematic study of the DOS spectrum and corresponding transport
characteristics of an increasingly disordered 2D electron system is needed to help
illuminate the crucial interplay between disorder and correlations as the system is
brought from the weakly to the strongly localized regime.  In the present Letter we
present such a study using ultra-thin Be films with low temperature sheet
resistances ranging from $R=500\Omega$ to $2.6M\Omega$.

	The theory of interaction effects in disordered electronic systems has for the
most part been developed in two extreme limits.  In the weak disorder/interaction 2D limit it is
known that the primary effect of $e-e$ interactions is to produce a logarithmic suppression of the
density of states (DOS) at the Fermi energy \cite{Altshuler}, $\delta N\sim
-ln(V)$.  This is commonly known as the zero bias anomaly (ZBA) and has been well established in
a number of different systems via tunneling measurements of the DOS
\cite{Imry,Gershenson,White}.  This depletion of the DOS is perturbative and results in a weakly
metallic 
$ln(T)$ transport conductivity \onlinecite{LeeRam}.  In the opposite limit, i.e. strongly
insulating regime, Efros and Shklovskii
\cite{ES1,ES2} have shown that the Coulombic
interactions can produce a {\em non-perturbative} gap in the DOS, which is commonly known as
the Coulomb gap.  The 3D Efros-Shklovskii Coulomb gap, which has a quadratic
energy dependence, has only recently been observed \cite{MLee1,MLee2}.  Interestingly, the
2D Coulomb gap is expected to be linear in energy \onlinecite{ES2},
\begin{equation} 
N(eV)=\frac{\alpha(4\pi\epsilon_{o}\kappa)^2|eV|}{e^4},
\end{equation}
where $\kappa$ is the relative dielectric constant, $\epsilon_o$ is the permittivity
of free space, and
$\alpha$ is a constant of order unity. Despite Eq.(1) having been in the literature
for more than 20 years now, there has been no direct spectroscopic verification of
its peculiar linear energy dependence due to the intrinsic technical
difficulties associated with measuring the DOS in thin insulating films \cite{Ashoori}.   
Nevertheless a direct measurement of this gap is important in that it is believed to play a
crucial role in the electron transport \cite{Valles} and thermoelectric \cite{Thermo}
properties of highly disordered 2D metals and semiconductors.  For instance, it is well known
that insulating films typically obey a modified variable range hopping law of the form
\begin{equation} 
R(T)=R_o\exp{(T_o/T)^\nu},
\end{equation}   
where R is the film sheet resistance, and $R_o$ is a constant. 
In the case of a flat DOS near the Fermi energy, the film simply obeys Mott's
variable range hopping law with $\nu=1/3$ \cite{Mott}.  If there is a simple gap in
the DOS then
$T_o$ is the gap energy for fixed range hopping and $\nu=1$.  Finally, if the DOS spectrum is
given by Eq.(1) then one expects
$\nu=1/2$, $R_o$ to be of the order of the quantum resistance $R_Q
=h/e^2$ \cite{ES3} and
\begin{equation} 
T_o=2.8e^2/(k_B4\pi\epsilon_o\kappa\xi),
\end{equation} 
where $k_B$ is the Boltzman constant and $\xi$ is the localization length
\onlinecite{ES2}.  Thus when the $\nu=1/2$ hopping form is observed, the DOS
spectrum is simply assumed to be that of Eq.(1).  In this Letter we examine this
assumption by presenting the first systematic electron tunneling study of the DOS in uniformly
disordered metal films whose transport properties range from that of weakly metallic to strongly
insulating.  We show that with increasing film resistance the DOS evolves from an essentially flat
spectrum with a perturbative $ln(V)$ ZBA to that of a Coulomb gap given by Eq.(1).  We also show
that the emergence of the Coulomb gap coincides with the emergence of the $\nu=1/2$ hopping
behavior of Eqs.(2) and (3).

The Be films used in the present study ranged in thickness from $1.5$ - $2.0 nm$
with corresponding sheet resistances
$R=500\Omega-3M\Omega$ at $T=50mK$. They were deposited by thermally evaporating $99.5\verb+%+$
pure beryllium powder onto fire polished glass substrates held at $84K$.  The
evaporations were made in a $4$x$10^{-7}$ Torr vacuum at a rate $\sim0.30 nm/s$.  The
film area was $1.5 mm$ x $4.5 mm$. Scanning force micrographs of the films' exposed oxide surface
did not reveal any salient morphological features down to the $0.7 nm$ resolution of the
instrument.  In fact, the films seemed to be as "smooth" as the fire-polished glass on which they
were deposited.  In addition, a transmission electron microstructural analysis of
$15nm$ thick Be films deposited on cleaved NaCl crystals at $84K$ revealed that the
films were composed of an ultra-fine base structure that was interspersed with
$5-15 nm$ Be nanocrystallites.  Electron diffraction measurements showed no
diffraction from the metallic base structure suggesting that it was amorphous. 
Similarly the oxide (BeO) produced a broad, continuous diffraction ring indicating
its grain size was $<1 nm$ \cite{Be}.   

	The tunnel junctions were formed by exposing the films to atmosphere for
0.1 - 3 hours in order to form a native oxide, then a $20nm$ thick Ag
counter-electrode was deposited directly on top of the film with the oxide serving
as the tunnel barrier.  The junction area was $0.7mm$ x $0.7mm$.  This technique
produced tunnel junction resistances of order $R_{TJ}\sim1k\Omega$ to $1000k\Omega$
depending upon the exposure time and other factors. We were always careful to
ensure that $R_{TJ}>>R_{film}$ at low temperatures. Our lowest resistance Be films
superconducted \onlinecite{Be} which allowed us to test the integrity of the junctions by
measuring the dc I-V characteristics at temperatures well below the superconducting transition
temperature $T_c$.  The sub-gap impedance of a "good" junction was always greater than
$10^8\Omega$ \cite{Tinkham}.   
  
	Shown in the inset of Fig.1 is the normal state conductance of the $2600\Omega$
film as a function of $ln(T)$.  This sample had a superconducting transition
temperature $T_c=0.33K$ which was suppressed by the application of a magnetic field
oriented along the film surface.  ($T_c$ was a monotonically
decreasing function of $R$.)  This sample had the expected $ln(T)$ weakly insulating
behavior with some rounding below $100mK$. In stark contrast, the main body of Fig. 1 shows the
activated-like behavior of the $2.6M\Omega$ film.  The solid line is a least squares fit to the
data.  The linearity of the data in Fig.1 shows unequivocally that the hopping exponent
$\nu=1/2$.  Furthermore the slope and intercept of the fit determine the parameters
$T_o\simeq1.6K$ and $R_o\simeq R_Q/2$ in Eq.(2).  It appears that $R_o$ is universal in that
it is of order the quantum resistance.  

	In principle we can also use Eq.(3) along with the measured value of $T_o$ to calculate a
localization length $\xi$.  However the  relative dielectric constant
$\kappa$ in Eq.(3) is unknown for a highly disordered metal film.  Alternatively, if
one instead assumes a reasonable value of $\xi\sim 1nm$ and takes $T_o\sim 1K$ then Eq.(3) predicts
$\kappa\sim 10^4$.  This value seems reasonable in that it lies between the  metallic and
insulating limits of $\kappa\sim\infty$ and
$\kappa\sim10$ respectively.  This is not an issue in 3D semiconducting systems where
the dielectric constant is well defined in the insulating phase.  In any case, it is
evident that Fig.1 is consistent with the existence of the Coulomb gap described by
Eq.(1).  Accordingly, if the Coulomb gap is indeed measurable in 2D, then it will be seen in a
sample such as that of Fig.1.
	
	Electron tunneling provides one of the most direct experimental measures of the DOS
spectrum of an electronic system.  The usual tunnel junction geometry consists an
electronically "inert" metallic counter-electrode which is separated from the film
by a highly insulating barrier, usually a few angstroms of oxide.  When a potential
is placed across the barrier electrons tunnel from the counter-electrode and into
the film.  At low temperatures the tunnel current is proportional to both the
counter-electrode's and the film's DOS \cite{Wolf}.  For the technique to be useful
in highly resistive films, however, care must be taken to insure that the impedance
of the tunnel junction is much higher than that of the film.  Otherwise,
some of the measured voltage drop will actually occur in the film itself, since
the tunnel current must be drained off through the film.  Additionally, there is a
legitimate concern that screening effects associated with the close proximity of
the highly conducting counter-electrode might "wash out" any Coulombic features in
the  DOS.  Fortunately, this has been shown not to be the case in recent measurements of
the Coulomb gap in 3D Si:B \onlinecite{MLee2}.

	In order to demonstrate the resolution of our tunnel junctions and the 2D nature of the Be films,
we have plotted in Fig. 2 the tunneling conductance of a $R=530\Omega$ superconducting film,
$T_c=0.55K$, in a parallel magnetic field just below its parallel critical field
$H_{c||}=1.2T$.  Note that we can resolve two peaks on either side of $V=0$.  These peaks
result from the Zeeman splitting of the usual BCS DOS peak and indicate that the
superconductivity was spin-paramagnetically limited
\cite{Fulde}.  The inset of Fig. 2 shows the normal state tunneling conductance of the same film
in a supercritical parallel field.  The normal state displays the expected $ln(V)$ DOS anomaly
which is consistent with the the $ln(T)$ transport behavior shown in the inset of Fig.1.

	In Fig.3 we show the evolution of the DOS with increasing disorder by plotting tunneling spectra
of Be films with $T=50mK$ sheet resistances of $R=530\Omega$, $2600\Omega$, $16000\Omega$, and
$2.6M\Omega$.  The
$530\Omega$ and $2600\Omega$ samples superconducted with $T_c=0.55K$ and
$T_c=0.33K$ respectively.  For those samples a parallel magnetic field was
applied in order to suppress the superconducting state and a standard ac lock-in
technique was used to measure tunneling conductance G directly.  In the higher resistance films,
the conductance was obtained by numerically differentiating dc I-V curves.  Note that
in each data set there is a significant ZBA.  In the $530\Omega$ and
$2600\Omega$ films the ZBA was of the weak disorder form  $\delta N/N\sim -ln(V)$.  However, as is
clearly evident in Fig. 3, the ZBA was no longer perturbative in films with
$R>10^4\Omega$.  An exponential growth in the ZBA is shown in Fig. 4 where we have plotted
$G(0)/G(15mV)$ as a function of R.  By making a linear fit to the data we find that $G(0)\sim
exp(-R/R_c)$ where $R_c=6000\Omega\approx R_Q/4$.

	The most interesting attribute of the data in Fig.3 is that the
conductance traces become more linear at the highest resistances studied. The energy dependence 
of the $16000\Omega$ curve represents the DOS spectrum 
of a film somewhere between the weak and strong disorder limits and is correspondingly neither
$ln(V)$ nor $|V|$ in form.  In fact, this spectrum is best described by a $V^{1/2}$ dependence. 
In contrast, the $2.6M\Omega$ spectrum is quite linear and symmetric.  For a consistency check,
we also measured the dc I-V characteristic of the films over the same current and temperature
ranges that were used in the tunneling measurements.  The nonlinearity in these transport I-V's
were negligible in comparison to the tunneling I-V's indicating that the traces in Fig.3 are
representative of the tunneling behavior. The solid line through the $2.6M\Omega$ data is a least
squares fit to the form $G(V)=\beta|V|$ where $\beta$ is an adjustable parameter.    We believe
that the linear behavior of this data represents the Efros-Shklovskii gap as given by Eq.(1).  
Unfortunately, we were unable to determine an absolute normalization of the tunneling
conductance.  The spectrum remained linear up to our maximum biases of $\sim100mV$.  Consequently
we cannot make a quantitative comparison between Eq.(1) and the tunneling data.  Nevertheless, the
observation of a linear spectrum in films that show the expected hopping transport behavior is
compelling.

In conclusion, we have made the first direct spectroscopic measurements of
the Efros-Shklovskii Coulomb gap in a 2D system.  We find the the gap emerges
from an exponential growth in the ZBA as the film sheet resistance is
increased.  The details of the activated transport behavior of films showing
the gap are in excellent agreement with theory and have a universal prefactor of the order of the
quantum resistance.  In principle, spectroscopic studies of
the Coulomb gap in films deposited {\it in situ} would allow for an absolute calibration of
the tunneling conductance since a variety of film thickness could be investigated using a
single tunnel junction.  Such measurements could then be used in conjunction with transport
measurements to extract the two primary microscopic parameters of the theory, $\kappa$ and $\xi$,
via Eqs.(1) and (3). 

We gratefully acknowledge discussions with Boris Shklovskii, Vladimir Dobrosavljevic, Boris
Altshuler.  This work was supported by the NSF under Grant No.'s DMR 9972151 and DMR 97-02690.


%

\begin{figure}
\caption{Semi-log plot of the resistance of the $2.6M\Omega$ Be film
as a function of $T^{-1/2}$.  The solid line is a linear fit to the data
for which we get $T_o=1.6K$ and $R_o\approx h/(2e^2)$, see Eq.(2).  Inset: 
Conductance of the $2600\Omega$ film as a function of $ln(T)$.  The solid line is a guide to the
eye.}
 \label{Figure 1}
 \end{figure}

\begin{figure}
 \caption{Tunnel conductance in the superconducting state of the $530\Omega$ film in a parallel
magnetic field $H_{||}=1.1T$. The critical parallel field was $H_{c||}=1.2T$.  The tunneling
spectrum demonstrates that we can resolve the Zeeman splitting of the usual BCS DOS.  Inset:
Normal state tunneling conductance of the same film as a function of $ln(V)$ at $H_{||}=2.5T$.}
 \label{Figure 2}
 \end{figure}

\begin{figure}
 \caption{Tunnel conductances normalized to G(15mV) for Be films with $T=50mK$
resistances of
$R=530\Omega, 2600\Omega, 16000\Omega$ and $2.6M\Omega$ (top to bottom).  The solid
lines are a best fit to the form $G(V)=\beta|V|$, where $\beta$ is an adjustable
parameter. The $2.6M\Omega$ film had a  $R_{TJ}=0.6M\Omega$ junction.  To insure that
$R_{film}<<R_{TJ}$ we took the $2.6M\Omega$ data at $700mK$, see Fig. 1.  The other curves were
measured at 50mK.}
 \label{Figure 3}
 \end{figure}

\begin{figure}
\caption{Semi-log plot of the normalized zero bias tunneling conductance as a
function of $R$.  The solid line is linear fit to the data.}
 \label{Figure 4}
 \end{figure}
%

%


\begin{references}
\bibitem[*] {Vladimir} Permanent address: Ioffe Physical Technical Institute (PTI), Russian
Academy of Sciences, Polytekhnicheskaya St., 26, 194021, St. Petersburg, Russia.
 
\bibitem{gang4} E. Abrahams, P.W. Anderson, D.C. Licciardello, and T.V.
Ramakrishnan, Phys. Rev. Lett. {\bf42}, 673 (1979).
\bibitem{LeeRam} P.A. Lee and T.V. Ramakrishnan, Rev. Mod. Phys. {\bf57}, 287
(1985).
\bibitem{Krav95} S.V. Kravchenko, W.E. Mason, G.E. Bowker, J.E. Furneaux, V.M.
Pudalov, and M. D'Iorio, Phys. Rev. B {\bf51}, 7038 (1995).
\bibitem{Sim97} D. Simonian, S.V. Kravachenko, M.P. Sarachik, and V.M. Pudalov,
Phys. Rev. Lett. {\bf79}, 2304 (1997).
\bibitem{Dobro} V. Dobrosavljevic, E. Abrahams, E. Miranda, and S. Chakravarty,
Phys. Rev. Lett. {\bf79}, 455 (1997).
\bibitem{Altshuler} B.L. Altshuler, A.G. Aronov, M.E. Gershenson, and Yu.V.
Sharvin, Sov. Sci. Rev. A. Phys. Vol. {\bf9}, 223 (1987).
\bibitem{Imry} Yoseph Imry and Zvi Ovadyahu, Phys. Rev. Lett. {\bf49}, 841
(1982).
\bibitem{Gershenson} M.E. Gershenson, V.N. Gubankov, M.I. Falei, Pis'ma Zh.
Eksp. Teor. Fiz. {\bf41}, 435 (1985) [JETP Lett. {\bf41}, 535 (1985)].
\bibitem{White} A.W. White, R.C. Dynes, and J.P. Garno, Phys. Rev. B {\bf31},
1174 (1985).
\bibitem{ES1} A.L. Efros and B.I. Shklovskii, J. Phys. C {\bf8}, L49 (1975).
\bibitem{ES2} B.I. Shklovskii and A.L. Efros, {\it Electronic Properties of Doped
Semiconductors}, (Springer, New York, 1984).

\bibitem{MLee1} J.G. Massey and Mark Lee, Phys. Rev. Lett. {\bf75}, 4266 (1995).

\bibitem{MLee2} J.G. Massey and Mark Lee, Phys. Rev. Lett. {\bf77}, 3399 (1996).

\bibitem{Ashoori} H.B. Chan, P.I. Glicofridis, R.C. Ashoori, and M.R. Melloch,
Phys. Rev. Lett. {\bf 79}, 2867 (1997).

\bibitem{Valles} Shih-Ying Hsu and J.M. Valles, Jr., Phys.
Rev. Lett. {\bf74}, 2331 (1995); Y. Shapir and Z. Ovadyahu, Phys. Rev. B {\bf40}, 12441 (1989);
Youzhu Zhang and M.P Sarachik, Phys. Rev. B {\bf43}, 7212 (1991).

\bibitem{Thermo} J.J. Lin, J.L. Cohn, and C. Uher in {\it Anderson Localization}, ed. by T. Ando
and H. Fukuyama, p. 186 (Springer-Verlag, New York, 1988).

\bibitem{Mott} N.F. Mott and G.A. Davis, {\it Electronic Processes in
Non-Crystalline Materials}, 2nd ed. (Clarendon, Oxford, 1979).
\bibitem{ES3} R.Berkovits and B.I.Shklovskii, J. Phys. Condens. Mat. {\bf11}, 779
(1998).
\bibitem{Be} P.W. Adams, P. Herron, and E.I. Meletis, Phys. Rev. B {\bf58}, 2952
(1998).
\bibitem{Tinkham} M. Tinkam, {\it Introduction to Superconductivity} (McGraw-Hill,
New York, 1996).
\bibitem{Wolf} E.L. Wolf, {\it Principles of Electron Tunneling Spectroscopy}
(Oxford Univ. Press, Oxford, 1985).
\bibitem{Fulde} P. Fulde, Adv. Phys. {\bf22}, 667 (1973).



\end{references}
\end{document}